\documentclass[pra,aps,twocolumn,10pt,showpacs,groupedaddress,superscriptaddress,floatfix,notitlepage]{revtex4-2}
\usepackage{amsmath,amsfonts,amssymb,graphics,graphicx,epsfig,color,times}
\usepackage[utf8x]{inputenc}
\usepackage{color}
\usepackage{bbm, dsfont} 
\usepackage{subfigure}
\usepackage{hyperref}
\usepackage{mathrsfs}
\usepackage{verbatim}
\usepackage{physics}

\begin{document}
\title{High-fidelity Rydberg controlled-\texorpdfstring{$Z$}{Z} gates with optimized pulses}

\author{T. H. Chang}
\affiliation{Institute of Atomic and Molecular Sciences, Academia Sinica, Taipei 10617, Taiwan.}
\author{T. N. Wang}
\affiliation{Institute of Atomic and Molecular Sciences, Academia Sinica, Taipei 10617, Taiwan.}
\affiliation{Graduate Institute of Applied physics, National Taiwan University, Taipei 10617, Taiwan}
\author{H. H. Jen}
 \email{sappyjen@gmail.com}
\affiliation{Institute of Atomic and Molecular Sciences, Academia Sinica, Taipei 10617, Taiwan.}
\affiliation{Physics Division, National Center for Theoretical Sciences, Taipei 10617, Taiwan}
\author{Y.-C. Chen}
 \email{chenyc@pub.iams.sinica.edu.tw}
\affiliation{Institute of Atomic and Molecular Sciences, Academia Sinica, Taipei 10617, Taiwan.}
 
\date{\today}
\renewcommand{\r}{\mathbf{r}}
\newcommand{\f}{\mathbf{f}}
\renewcommand{\k}{\mathbf{k}}
\def\p{\mathbf{p}}
\def\q{\mathbf{q}}
\def\bea{\begin{eqnarray}}
\def\eea{\end{eqnarray}}
\def\ba{\begin{array}}
\def\ea{\end{array}}
\def\bdm{\begin{displaymath}}
\def\edm{\end{displaymath}}
\def\red{\color{red}}
\begin{abstract}
High-fidelity controlled-$Z$ ($C_Z$) gates are essential and mandatory to build a large-scale quantum computer. In neutral atoms, the strong dipole-dipole interactions between their Rydberg states make them one of the pioneering platforms to implement $C_Z$ gates. Here we numerically investigate the optimized pulses to generate a high-fidelity Rydberg $C_{Z}$ gate in a three-level ladder-type atomic system. By tuning the temporal shapes of Gaussian or segmented pulses, the populations on the intermediate excited states are shown to be suppressed within the symmetric gate operation protocol, which leads to a $C_{Z}$ gate with a high Bell fidelity up to $99.92\%$. These optimized pulses are robust to thermal fluctuations and the excitation field variations. Our results promise a high-fidelity and fast gate operation under amenable and controllable experimental parameters, which goes beyond the adiabatic operation regime under a finite Blockade strength.  
\end{abstract}
\maketitle

\section{Introduction} 

Single-qubit manipulations and two-qubit conditional operations are the building blocks as universal quantum gates in quantum computation \cite{DiVincenzo2000, Morgado2021}. Based on the neutral-atom platform with Rydberg interactions \cite{Jaksch2000, Saffman2010, Shi2022}, an entangling $C_Z$ gate can be feasible \cite{Urban2009, Gaetan2009}, in addition to single-qubit gates facilitated by individual laser operations between two hyperfine ground states. These universal quantum gates are the foundations to generate multiqubit entanglement \cite{Graham2022}, which along with coherent and parallel operations can further allow scalable quantum processing \cite{Bluvstein2022}. With the capability of connectivity \cite{Ramette2022}, error-correcting quantum computations \cite{GoogleAI2021, Cong2022} under a small error threshold can be envisioned in the near future. While single-qubit gates can achieve a higher fidelity, an improved and refined fidelity of two-qubit gates is relatively difficult and is essential to realize a large-scale quantum computer. In other words, this high-fidelity two-qubit gate sets the fundamental limit in deep quantum circuits \cite{Preskill2018} or scalable quantum computation \cite{Devitt2013}. 

There has been many efforts in optimizing the two-qubit gate operation protocol in neutral atoms. To characterize the performance of $C_Z$ gates, the fidelity of Bell state preparation can be applied as a quantification measure \cite{Sackett2000}. Two commonly acceptable criterions to achieve a high fidelity in generating Bell states involve significant Rydberg interactions and minimal spontaneous decays. These result in, respectively, a low blockade leakage owing to finite excitations to two Rydberg states and negligible irreversible effect owing to dissipations and decoherences. Under the finite Rydberg interaction strength and lifetime, there is still a lot of room for improvement, for example, by using shaped pulses \cite{Goerz2011, Goerz2014, Theis2016}, a two-atom dark state \cite{Petrosyan2017}, spin-echo technique \cite{Shi2018}, quantum interference of laser excitations \cite{Shi2019}, dual-rail Rydberg configurations \cite{Shi2020}, and symmetric gates \cite{Levine2019, Saffman2020, Fu2022, Jandura2022, Pagano2022, Li2022, Evered2023}. 

The symmetric \cite{Levine2019} or global \cite{Jandura2022, Fromonteil2022, Evered2023} $C_Z$ gates have been shown to have a high-fidelity performance as well as a faster gate time compared to the originally proposed individual $\pi$-$2\pi$-$\pi$ pulse sequences on the respective control-target-control qubits \cite{Jaksch2000, Saffman2010}. The enhanced performance results from the dynamical phase compensations in the symmetric gate protocols with an optimized laser field detuning \cite{Levine2019}, which can be further improved by optimizing pulse shapes \cite{Jandura2022, Pagano2022, Evered2023}. This shows the capability of controllable laser fields with tunable amplitudes and detunings, and the opportunities provided by these optimized laser pulses to achieve an even higher fidelity two-qubit gates \cite{Mohan2022}. Meanwhile, a practical experimental implementation \cite{Fu2022} prefers a scheme with straightforward and tractable operating parameters, which can hugely reduce the complexity and the uncontrollable systematic errors in experiments. In contrast to a purely theoretical study often considering an idealistic system setup, an optimization protocol in the symmetric $C_Z$ gate operation under a realistic experimental consideration with an intermediate excited state is less explored. Furthermore, optimized protocols that go beyond adiabatic passages \cite{Rao2014, Beterov2016, Mitra2020, Beterov2020, Saffman2020, Li2021, He2022, Mitra2023} are not fully investigated, which can provide new parameter regimes for an improved gate fidelity and can be potentially suitable for near-term implementations.  

Here we consider a three-level ladder-type alkali atomic system with two atoms trapped individually in optical tweezers \cite{Endres2016, Barredo2016, Barredo2018, Schlosser2023}. With finite Rydberg interactions, we numerically investigate the optimized pulses to generate a high-fidelity Rydberg $C_{Z}$ gate in the symmetric gate operation protocol. Within a broad range of pulse duration under tunable temporal Gaussian or segmented shapes, the optimization protocol presents a suppressed intermediate excited-state population, which gives rise a $C_{Z}$ gate with a high Bell fidelity up to $0.9992$. These optimized pulses are robust to thermal fluctuations but are relatively more sensitive to the excitation field variations. Our results provide a promising protocol for an improved high-fidelity and fast gate operation under practical experimental parameters, which explores a parameter regime beyond the adiabatic operation and paves the way toward deeper quantum circuits and foreseeable fault-tolerant quantum computing \cite{GoogleAI2021, Cong2022, Svore2007, Fowler2012, Lai2014, Iyer2022, Cleland2022, Liou2023, GoogleAI2023}.  

The remainder of the paper is organized as follows. In Sec. II, we introduce the Hamiltonian that governs the Rydberg interactions between two trapped atoms. In Sec. III, we introduce the optimization protocol and present the optimized results of $C_Z$ gates with optimized pulses in either Gaussian or segmented shapes. In Sec. IV, we explore the effect of thermal and excitation field variations on the performance of the $C_Z$ gates. We discuss and conclude in Sec. V.  

\section{Theoretical model}

\begin{figure}[t]
    \includegraphics[width=0.49\textwidth]{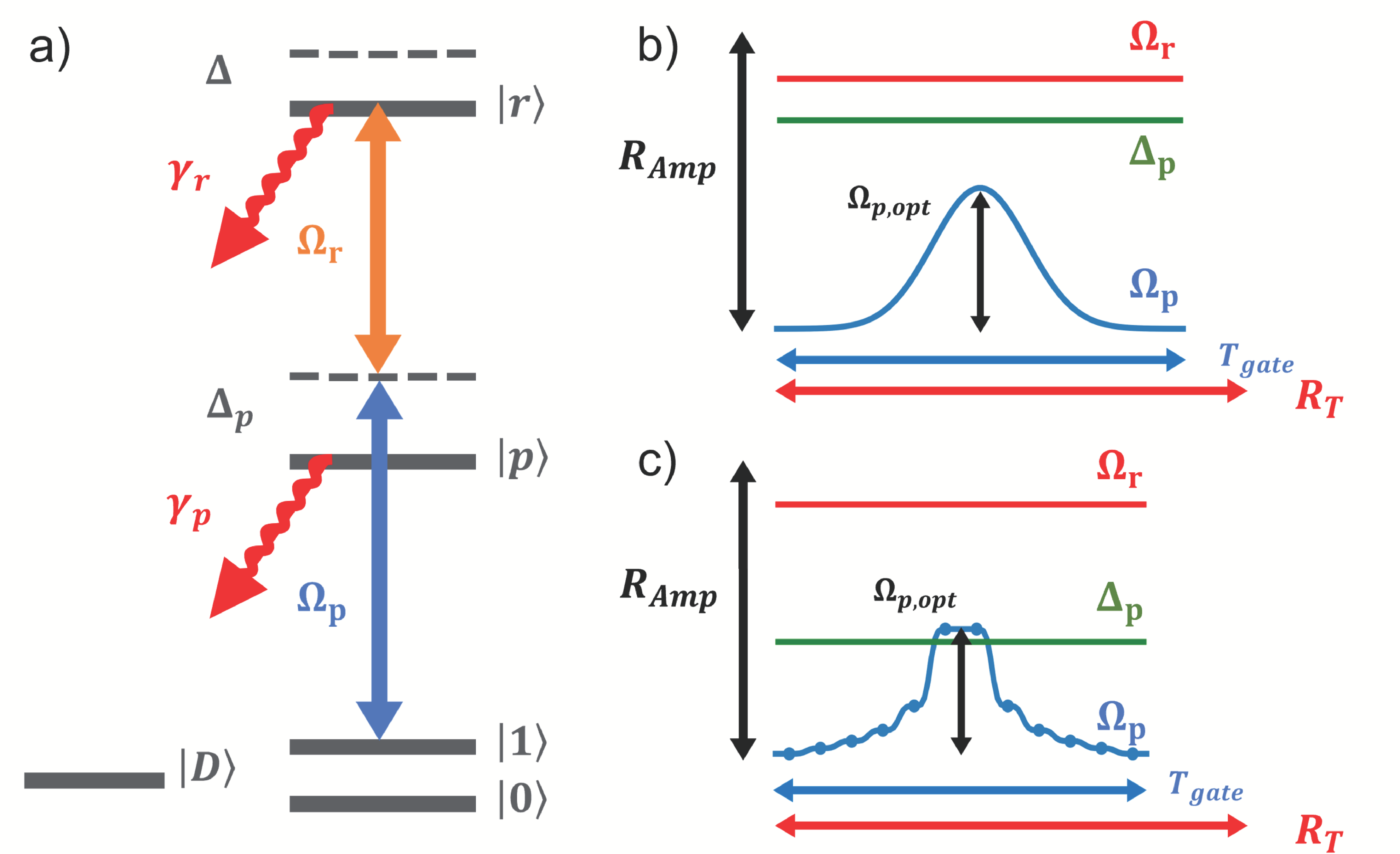}
    \caption{Optimized pulse scheme in $C_Z$ gate. a) Three-level ladder type atomic system with the Rydberg state $|r\rangle$, intermediate excited state $|p\rangle$, physical qubit states $|0\rangle$ and $|1\rangle$, and dark trapped state $|D\rangle$. We apply two global laser excitation fields $\Omega_p$ and $\Omega_r$ to fulfill the $C_Z$ gate operation by taking into account the various decay channels $\gamma_r$ and $\gamma_p$ under a laser detuning $\Delta_p$. Two-photon resonance at $\Delta=0$ is often assumed for efficient gate operation using Rydberg blockade. We consider two optimization schemes in the laser fields $\Omega_p$ with a Gaussian wave form or segmented pulse in b) and c), respectively. Optimization parameters involve $\Omega_p$ and $T_{\rm gate}$, while $R_{\rm Amp}$ and $R_T$ are the allowed ranges for these parameters. Other laser parameters of field amplitudes $\Omega_r$ and detuning $\Delta_p$ are kept constant in time but tunable and optimized as well under the same optimization regions for $\Omega_p$.}\label{fig1}
\end{figure}

We consider two atoms with a ladder-type three-level atomic structure as shown in Fig. \ref{fig1}(a), where the Rydberg states can be excited homogeneously via a two-photon excitation through an intermediate excited state \cite{Saffman2020, Evered2023}. The computational basis can be encoded in the hyperfine ground states $\ket{0}$ and $\ket{1}$, where $\ket{1}$ is coupled to the intermediate state $\ket{p}$ with a Rabi frequency $\Omega_p$ and detuning $\Delta_p$, and $\ket{p}$ is further excited to the Rydberg state $\ket{r}$ with a Rabi frequency $\Omega_r$ and detuning $\Delta_r$. We utilize this symmetric setup with global driving fields to achieve the goal of $C_Z$ gate which transforms $\{\ket{00}, \ket{01}, \ket{10}, \ket{11} \}$ to $\{\ket{00}, \ket{01}, \ket{10}, -\ket{11} \}$, equivalent to a projection operation of $2|00\rangle\langle 00
|-I$ up to a single qubit rotation \cite{Levine2019}. This controlled-phase gate has been proposed by stimulated Raman adiabatic passage pulses \cite{Moller2008}, where the initial states of $\ket{01}$ and $\ket{10}$ will follow an energy dark state during the gate operation and accumulate no dynamical phases under the two-photon resonance condition $\Delta=\Delta_p+\Delta_r=0$. Meanwhile, with a large detuned first laser field $\Delta_{p}\gg\Omega_{p}, \Omega_{r}$, the initial state of $\ket{11}$ will follow a $2\pi$ rotation as $\ket{11} \rightarrow(\ket{1r}+\ket{r1})/\sqrt{2} \rightarrow\ket{11}$, which accumulates a required $\pi$ phase.

To investigate the performance of the $C_Z$ gate operation, we numerically solve the two-atom master equation in Lindblad forms, 
\bea
\frac{d\rho}{dt}=i\left[{\cal H, \rho}\right]+\Gamma[\rho],\label{rho}
\eea
from which we are able to obtain the system evolutions of two-atom density matrix $\rho$ for all four possible initial states in the computational bases. The interaction Hamiltonian can be written as 
\bea
{\cal H}=&&{\cal H_{\text{c}}} \otimes I
+ I\otimes{\cal H_{\text{t}}}
+ B_{rr}\ket{rr}\bra{rr},\\
{\cal H}_{i=\rm{c,t}}=&&\Delta_p(t)\ket{p}_i\bra{p} + \Delta(t)\ket{r}_i\bra{r} \nonumber\\
&&+ \left[\frac{\Omega_p(t)}{2}\ket{p}_i\bra{1} + \frac{\Omega_r(t)}{2}\ket{r}_i\bra{p} 
+ {\rm h.c.}\right],
\eea
where `c' and `t' stand for the control and the target qubit, respectively, $B_{rr}$ represents the strength of the Rydberg-Rydberg interaction between two Rydberg states, and $\Delta = \Delta_p + \Delta_r$ indicates the two-photon detuning. The Lindblad forms considering various decay channels in Fig. \ref{fig1}(a) can be expressed as 
\begin{eqnarray}
\Gamma[\rho]=
\sum_{i=c, t} 
\sum_{\substack{j=0, 1,\\D,p\delta_{k,r}}}
\sum_{k=p, r}
\Gamma^{i}_{jk}\rho\Gamma^{i^{\dagger}}_{jk}
- \Gamma^{i^{\dagger}}_{jk}\Gamma^{i}_{jk}\frac{\rho}{2}
- \frac{\rho}{2}\Gamma^{i^{\dagger}}_{j}\Gamma^{i}_{jk},\nonumber\\
\end{eqnarray}
where $\delta$ is the Kronecker delta function and $\Gamma^{i}_{jk}\equiv\sqrt{c_{jk}\gamma_{k}}\ket{j}_i\ket{k}$ with the specified decay ratios of $c_{0p}=0.1354$, $c_{1p}=0.2504$, $c_{Dp}=0.6142$, $c_{0r}=0.053$, $c_{1r}=0.053$, and $c_{Dr}=0.894$, based on $|p\rangle=|6P_{3/2}\rangle$ and $|r\rangle=|53S_{1/2}\rangle$ for rubidium atoms \cite{Levine2018, Evered2023}. The dark trapped state $\ket{D}$ is decoupled from the other atomic states, which can be occupied by the spontaneous decays from the excited states $\ket{p}$ and $\ket{r}$, where $\gamma_{r} = 1/(88~\mu \text{s})$ for the atoms at room temperature and $\gamma_{p} = 1/(0.11~\mu \text{s})$. We take these system parameters as an exemplary reference and comparison throughout the paper.

There should be two dominant error sources in the gate operation, which are the spontaneous decay and the Rydberg blockade error. The former is indispensable, especially in the two-photon excitation process through an intermediate excited state whose intrinsic lifetime is much shorter than the one in the Rydberg state, and this sets one of the limits in the gate performance owing to the unavoidable effect of decoherence. The latter, on the other hand, represents a finite Rydberg-Rydberg interaction and breaks the ideal assumption of infinite Blockade and associated perfect gate operation owing to the leakage to a finite Rydberg excitation, which again experiences the effect of dissipations in the Rydberg state. We note that, however, the effective lifetime in the intermediate state can be longer if a larger laser detuning is applied. Naively one would consider an even shorter excitation pulse along with significant Rydberg interaction, which would further counteract the effect of the intermediate excited-state decoherence and Rydberg state leakage to reduce the gate errors. However, this neglects the facts that a higher-level Rydberg state with more significant interaction goes with a more closely-spaced Rydberg energy levels, where nearby Rydberg state excitations become inevitable. This leakage turns to be even worse in the short-pulse excitation when its broadband spectrum exceeds the driving field detuning. Therefore, the quantum gate operation protocol becomes an optimal control problem which seems complicated, and multiple optimal solutions should exist when more controllable system parameters are involved. One of the approaches to achieve high enough $C_Z$ gate suggests using optimal pulses through pulse shaping with optimization algorithms. 

Here we adopt the method of differential evolution (DE) \cite{Storn1997, Zahedinejad2014}, which is a global optimization algorithm over continuous space for quantum control. It consists of three steps in each iteration of the protocol, including mutation, crossover, and selection. First, we create a population for the parameters we want to optimize. An individual in the population is a $D$-dimensional vector $X_{i}$ with $i=1,2,3,…,N_\text{P}$, where $D$ is the number of parameters and $N_\text{P}$ is the population size. For each iteration, four different individuals $X_{i}$, $X_{s1}$, $X_{s2}$, $X_{s3}$ will be sampled randomly from the population. The mutant vector $V_{i}$ is generated by
\begin{equation}
    V_{i} = X_{s1} + \mu (X_{s2} - X_{s3} )
\end{equation}
with $\mu \in [0,2]$ the mutation factor which controls the differential variation $X_{s2} - X_{s3}$.
Crossover is designed to increase the diversity of the perturbed parameter vectors, where the trail vector $U_{i}$ is obtained by
\begin{equation}
    U_{i} = 
    \begin{cases}
        V_{i} & \text{if}\ c(0,1) < \xi,\\
        X_{i} & \text{otherwise},
    \end{cases}
\end{equation}
with $c(0,1)$ the uniform random number between $0$ and $1$, sampled in every iteration, and the crossover rate $\xi \in (0,1)$. Finally, The member of the next generation $X'_{i}$ will be determined in the selection process 
\begin{equation}
    X'_{i} = 
    \begin{cases}
        U_{i} & \text{if}\ f(U_{i}) < f(X_{i}),\\
        X_{i} & \text{otherwise},
    \end{cases}
\end{equation}
where $f$ is the cost function of the optimization problem, which is the infidelity in our case. We note that $c(0,1)$ is different for each component of the $D$-dimensional vector, and we choose the population size as $N_p$ $=$ (the number of control parameters)$\times 15$ with $300$k generations used for each optimization. For the convergence criteria, we stop the algorithm when the fourth digit of the fidelity does not evolve, which is sufficient for us to identify its maximum under our considered physical parameter regimes. For higher digits in the fidelity, other corrections, for example, from the far-detuned Rydberg levels would be relevant, which is beyond the scope of the Hamiltonian in this work. 

We then apply DE method in optimizing Gaussian wave forms and segmented pulses with an extra free parameter of their pulse durations $T_{\rm gate}$, as shown schematically in Figs. \ref{fig1}(b) and \ref{fig1}(c) respectively. The reason why we focus on these two pulse regimes is that Gaussian wave form is common and adaptable in experiments, while the segmented pulses provide extra free parameters in time segments which should be more robust to system fluctuations. This optimized pulse procedure gives rise a fast and high-fidelity $C_Z$ gate performance, leading to a high Bell fidelity more than $0.999$. In the following two subsections, we focus on optimizing Gaussian wave forms and segmented pulses of the first laser field $\Omega_p$ while keeping the other laser parameters constant in time but tunable in our optimization procedure. 
      
\section{\texorpdfstring{$C_Z$}{CZ} gate with optimized pulses}

\subsection{Optimized Gaussian pulses}

First we focus on using Gaussian pulses to optimize $C_Z$ gates in the global driving scheme. Optimization parameters involve the maximum amplitude $\Omega_{p,\text{max}}$ and the total gate operation time $T_\text{gate}$ of the Gaussian profile $\Omega_p(t) = \Omega_{p,\text{max}}[e^{-(t-t_c)^2 / \tau^2} - a] / (1-a)$, where $t_c=T_{\rm gate}/2$ indicates the center of the pulse, an associated pulse width $\tau/T_\text{\rm gate} = 0.165$, and the offset $a = e^{-t_c^2 / \tau^2}$ is to give a zero amplitude at the beginning and the end of the pulse. We utilize DE approach to locate the optimized solution in these parameters within the designated ranges of field amplitudes $R_\text{Amp}$ and pulse durations $R_T$, respectively. We further keep $\Omega_r(t) = \Omega_{r,\text{max}}$ and $-\Delta_p(t) = \Delta_{p,\text{max}}$ constant in time but tunable within the same optimization ranges determined by $R_{\rm AMP}$. We use QuTip package \cite{Johansson2012, Johansson2013} to solve $\rho$ of the master equation in Eq. (\ref{rho}) and use Bell fidelity $F_{\rm Bell}\equiv (\rho_{0000}+\rho_{1111})/2+|\rho_{1010}|$ \cite{Sackett2000, Saffman2020} as an optimization quantity in DE approach. 

\begin{figure}[tb]
    \includegraphics[width=0.49\textwidth]{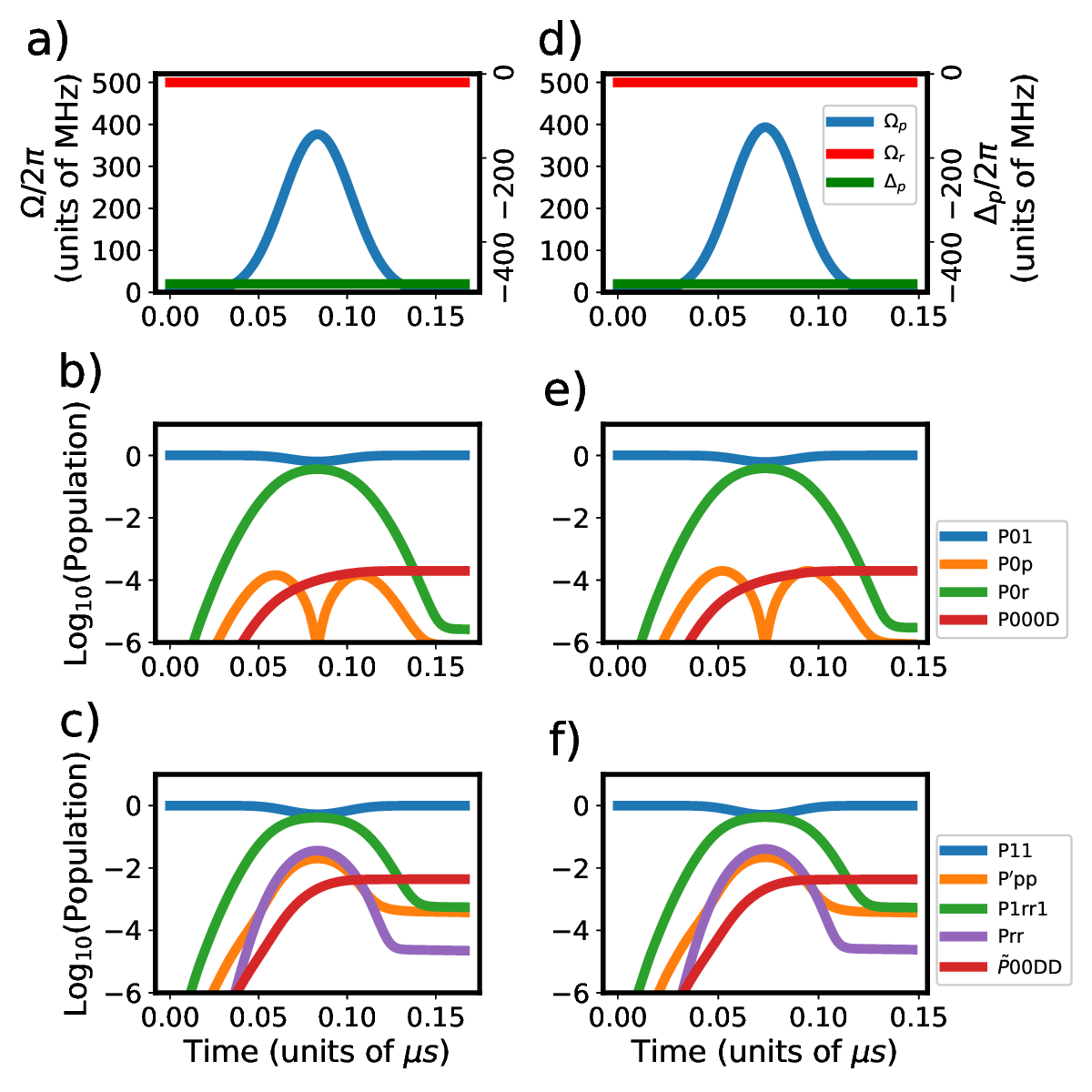}
    \caption{Numerical simulations of optimized Gaussian pulses. a) The optimized Gaussian pulse with Bell fidelity $F_{\rm Bell}\approx 0.9985$ under the blockade strength $B_{rr}/(2\pi)=250$ MHz. The allowed range of the pulse amplitudes $\Omega_{p(r),{\rm max}}/(2\pi)$ and laser detuning $\Delta_{p,{\rm max}}/(2\pi)$ is chosen as $R_{\rm AMP}=[0, 500]$ MHz and the allowed range for the total gate operation time is $R_T=[0, 10]$ $\mu$s with $\gamma_p=1/(0.11~\mu \text{s})$ and $\gamma_r=1/(88~\mu \text{s})$. b) The time evolution of various atomic populations for $\ket{01}$, $\ket{0p}$, $\ket{0r}$, and $\text{P}_{000D} = \text{P}_{00} + \text{P}_{0D}$ for the states of $\ket{00}$ and $\ket{0D}$ during the gate operation with the initial state $\ket{01}$ of the system. c) The time evolutions of the atomic populations for $\ket{11}$, $\ket{pp}$, $(\ket{1r}+\ket{r1})/\sqrt{2}$, $\ket{rr}$, and $\tilde{\text{P}}_{00DD}=\text{P}'_{00}+\text{P}'_{DD}-({\rm P}_{0D}+{\rm P}_{D0})$ for either of the system is in the state of $\ket{0}$ or $\ket{D}$ with the system's initial state $\ket{11}$. We define ${\rm P}'_{ii} = \sum_{j = 0, 1, D, p, r} { {\rm P}_{ij} + {\rm P}_{ji} - {\rm P}_{ii}}$ as the state populations in $|i\rangle$ for both atoms. The corresponding results of d,e,f) to a,b,c) are obtained under a different lifetime $\gamma_r=1/(44~\mu \text{s})$ as a comparison, which gives $F_{\rm Bell}\approx 0.9984$.}\label{fig1_500}
\end{figure}

As shown in Fig. \ref{fig1_500}, we take an example of $R_\text{Amp} = [0, 500]$ MHz and $R_T = [0, 10]~ \mu$s where we find $F_\text{Bell} \approx 0.9985$. We have chosen the two-photon resonance condition $\Delta=0$ which is indeed the optimized case and has also been confirmed in our numerical simulations. This performance of $C_Z$ gate is close to the result of $F_\text{Bell}=0.997$ under the scheme of stimulated rapid adiabatic passage (STIRAP) pulses \cite{Saffman2020} when $\Delta_p$ is much larger than $\Omega_p$ and $\Omega_r$. However, here we obtain the results in a relatively weak blockade regime, where in Fig. \ref{fig1_500}(a), the dominant state evolutions mainly involve the subspace of $|11\rangle$ and $(|1r\rangle+|r1\rangle)/\sqrt{2}$ as shown in Fig. \ref{fig1_500}(c) with comparable state populations in $|p\rangle$ and $|r\rangle$. From Figs. \ref{fig1_500}(b) and \ref{fig1_500}(c), we can attribute the main errors in global driving scheme to finite populations in the trapped dark state $|D\rangle$ that can never be redistributed and a leakage to the Rydberg state and the intermediate excited state $|p\rangle$, where decoherences indispensably compromise the gate fidelity owing to their finite lifetimes. As a comparison, we numerical simulate the corresponding results under a smaller lifetime $1/\gamma_r$ in Figs. \ref{fig1_500}(d-f) compared to Figs. \ref{fig1_500}(a-c). We obtain a similar high $F_\text{Bell}$, which can still be sustained in the optimization approach. This suggests that the harmful effect of short Rydberg state lifetime on the gate performance can be reduced under the optimized excitation pulses with a slightly higher intensity in $\Omega_p$. 

\begin{figure}[tb]
    \includegraphics[width=0.49\textwidth]{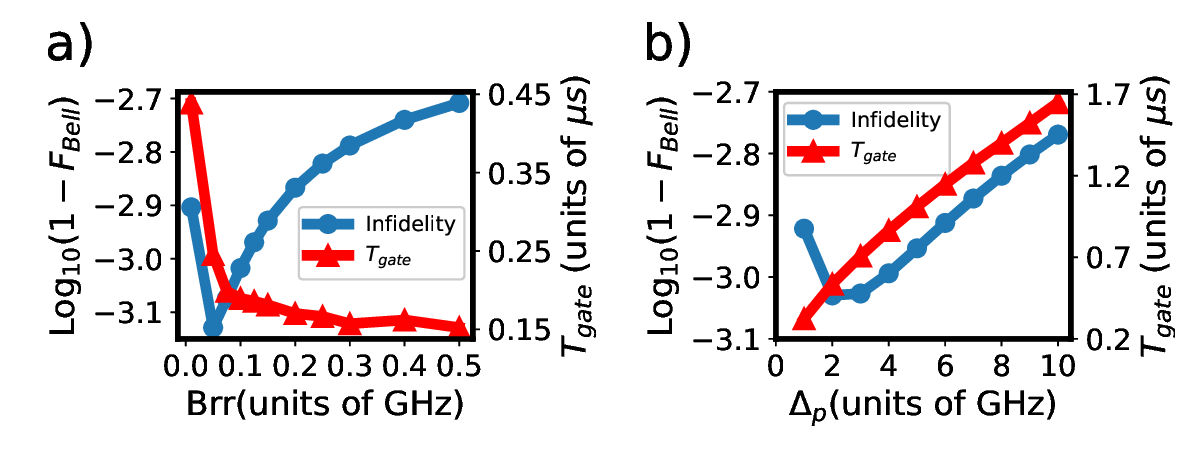}
    \caption{Bell infidelity and optimized gate operation time $T_{\rm gate}$ of Gaussian pulses. a) Within $R_{\rm AMP}=[0, 500]$ MHz, we optimize the pulse amplitudes $\Omega_{p(r),{\rm max}}/(2\pi)$ and laser detuning $\Delta_{p,{\rm max}}/(2\pi)$ for different $B_{rr}$, as well as the gate optimization time $T_{\rm gate}$ within $R_T = [0, 10]~ \mu$s. The results demonstrate the optimized high-fidelity pulse under weak blockade regime, where $F_\text{Bell} \approx 0.99926$ at $B_{rr}/(2\pi)=50$ MHz. b) With the same parameter range as in a), we fix $B_{rr}/(2\pi)=500$ MHz and explore the strong blockade regime at different $\Delta_p/(2\pi)$. There is a trade-off between the large detuned laser field and the gate operation time $T_{\rm gate}$, which gives the optimal $\Delta_p/(2\pi)=-2$ GHz with $F_\text{Bell} \approx 0.99907$.}\label{fig2}
\end{figure}

To further improve the fidelity of the $C_Z$ gate, we vary $B_{rr}$ or $\Delta_p$ with a fixed $B_{rr}$ to explore a broader parameter region, from which we expect that the optimized scheme would allow faster pulses with an outperformed gate fidelity owing to the mitigated effect of spontaneous decays. In Fig. \ref{fig2}, we plot the Bell infidelity instead for a better demonstration of the gate performance along with the optimized pulse duration $T_{\rm gate}$. As the Blockade strength $B_{rr}$ decreases in Fig. \ref{fig2}(a), we find that the new optimized schemes emerge with an improved gate fidelity. However, the fidelity becomes worse with an even weaker $B_{rr}$ owing to its associated longer gate time that causes decoherences. This worse performance can be attributed to the dissipations from the finite populations in the Rydberg states. We note that the theoretical model would break down if there is significant population in the doubly-excited Rydberg states, since F\"{o}rster resonances would be non-negligible \cite{Saffman2010} and lead to new error sources from unwanted energy transfer. The results in Fig. \ref{fig2}(a) indicate a nonadiabatic regime under a finite blockade strength in the optimization method, which we will elaborate more below. By contrast, in Fig. \ref{fig2}(b) under the strong blockade regime instead for a large $|\Delta_p|$, we again can identify the optimized parameter that leads to a better fidelity. As expected, a larger $|\Delta_p|$ requires a longer gate time to fulfill the $C_Z$ gates, which leads to the errors from the limited lifetime of the intermediate states. This error also becomes significant when a near-resonant laser field is applied. 

\begin{figure}[tb]
    \includegraphics[width=0.49\textwidth]{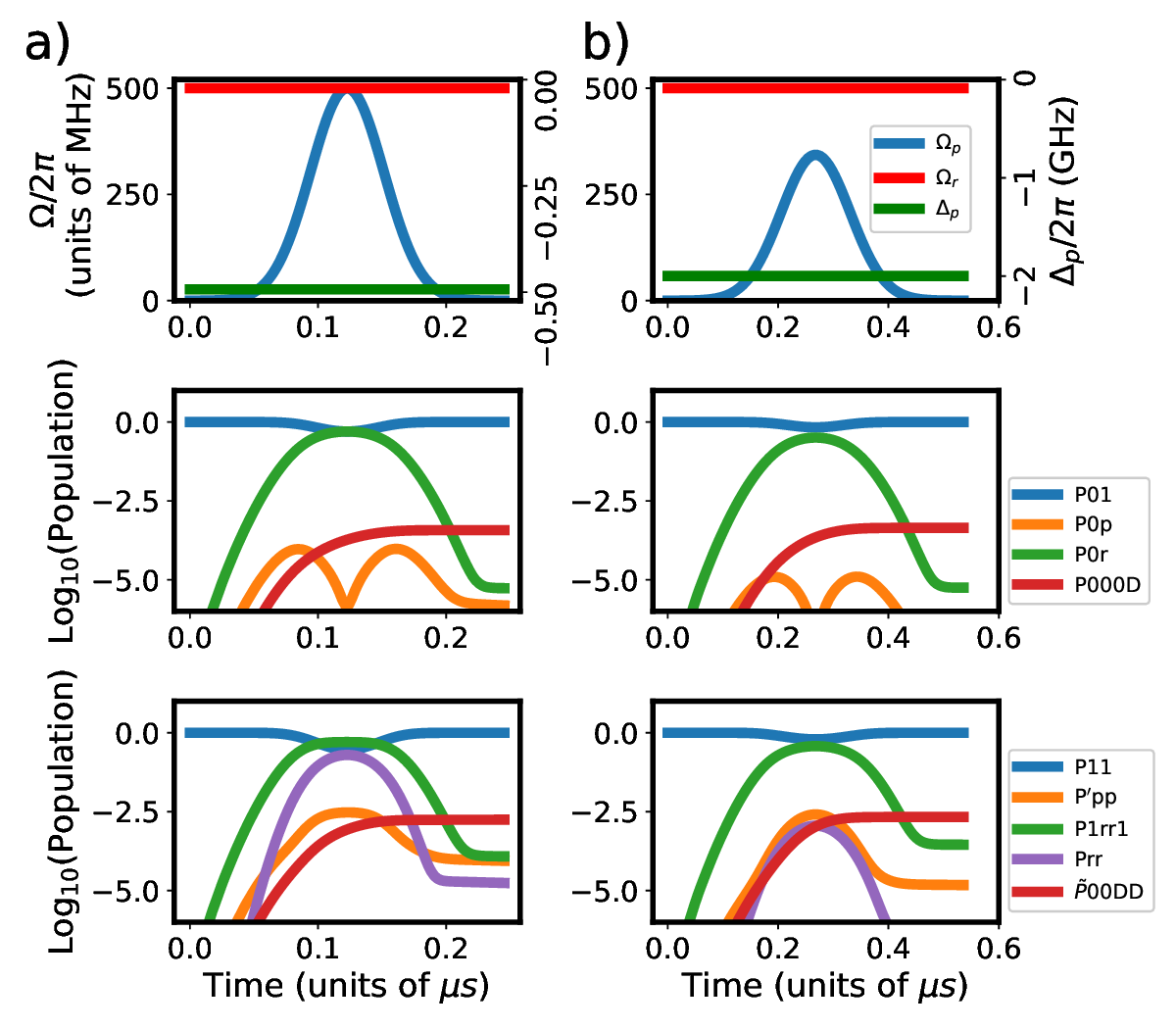}
    \caption{Optimized Gaussian pulses in Figs. \ref{fig2}(a) and \ref{fig2}(b) respectively for a) $B_{rr}/(2\pi)=50$ MHz and b) $\Delta_p/(2\pi)=-2$ GHz. We present in the top plots for the optimized pulse shape and laser parameters, in the middle and the bottom plots for the time evolutions of the atomic populations from the initial states $\ket {01}$ and $\ket{11}$, respectively. The fidelities are $F_{\rm Bell} \approx 0.99926$ and $0.99907$, respectively.}\label{fig3}
\end{figure}

In Fig. \ref{fig2}(a), we note that the trend of a less improving gate fidelity goes with a larger $B_{rr}$, which is contrary to the trend in the strong blockade regime often considered \cite{Saffman2010}. This is due to the weak blockade regime identified and optimized here when $B_{rr}/(2\pi)\ll 500$MHz, where both pulse amplitudes are comparable to the laser detuning. In this nonadiabatic regime, the populations in the intermediate states $|p\rangle$ are suppressed with rising but low populations in the doubly-excited Rydberg states [shown later in Fig. \ref{fig3}(a)], which leads to less errors from the decays of the intermediate state and as well as less errors from the dark trapped states owing to smaller leakages from the intermediate states. This is in contrast to the parameter regime close to the strong blockade at $B_{rr}/(2\pi)\gtrsim 500$MHz, where doubly-excited Rydberg populations are suppressed along with relatively significant populations in the intermediate state. Therefore, the reason why it seems unconventional in Fig. \ref{fig2}(a) that the fidelity is less as the blockade interaction increases is because we are sweeping the optimized system parameters from a weak to a relatively moderate blockade regime. For even larger $B_{rr}\gg 500$MHz, we find that the fidelity saturates without further improvement as $B_{rr}$ increases. This is due to the dark trapped state we introduce to account for any unwanted leaking channels, which posits the ultimate limit for gate performances. If there is no such leaking channel of the dark trapped state, the gate performance will behave as in the strong blockade regime. We note, however, that this leaking channel is indispensable in experimental implementations.

In this weak blockade regime, the gate fidelity will strongly depend on the fluctuations of $B_{rr}$. We take the optimized parameters in Fig. \ref{fig2}(a) and study the effect of variations of $B_{rr}$ in the gate fidelity. At $B_{rr}/(2\pi)=50$MHz, a $\pm 5\%$ deviation leads to a drop of $F_{\rm Bell}\approx 0.9992$ down to around $0.9968$. By contrast at $B_{rr}/(2\pi)=500$MHz, the same deviation leads to a drop of $F_{\rm Bell}\approx 0.9981$ down to around $0.9978$. This shows the fragility of the optimized weak blockade regime and demands a more precise control of the stability of atomic positions.  

We now take a closer look into these optimized schemes in Figs. \ref{fig2}(a) and \ref{fig2}(b), and plot them respectively in Figs. \ref{fig3}(a) and \ref{fig3}(b) at their optimal parameters. The former optimized scheme represents a nonadiabatic regime where relatively large laser fields of $\Omega_p$ and $\Omega_r$ are applied compared to the detuning $|\Delta_p|$, in contrast to the strong blockade regime in the latter scheme satisfying the adiabatic condition under a larger laser detuning than both laser field strengths. In the nonadiabatic regime, the optimized $\Omega_r$ and the peak value of $\Omega_p$ hit the optimization boundary, and the $C_Z$ gate operation is no longer under a significant Rydberg blockade regime. Therefore, we attribute this regime as a weak blockade regime, which goes beyond the adiabatic passage assumption and permits fast gates with $T_\text{gate}$ achieving sub-microsecond. This results in significantly reduced errors of the spontaneous decay from both the intermediate excited and Rydberg states, which gives rise to $F_\text{Bell}$ approaching $0.9992$. We note that in Fig. \ref{fig3}(a), the optimized laser detuning $\Delta_p$ and Rabi frequency $\Omega_r$ are both located at the optimization range boundary, as well as for the optimized $\Omega_r$ in Fig. \ref{fig3}(b). This manifests that the ultimate performance in $C_Z$ gates is limited by the optimization range for the system parameters, either in the adiabatic or the nonadiabatic regimes, and an even better performance is expected when the limits of the optimization protocols can be further released.  

Figure \ref{fig3} also demonstrates the suppressed populations in $|p\rangle$, and the main error is dominated by the population in the dark state $|D\rangle$ introduced in the theoretical model \cite{Saffman2020}. In our optimized global $C_Z$ gates, the errors from $|p\rangle$ are kept low in both adiabatic and nonadiabatic regimes, whereas a significant improvement emerges when $\Omega_p$ and $\Omega_r$ are made larger with shorter pulse durations. This optimization scheme provides not only high-fidelity $C_Z$ gates, but also fast gates under the optimized consideration, which can be feasible beyond the adiabatic regime. Next we further release the restriction of Gaussian forms and explore the optimization possibilities using segmented pulses. 

\subsection{Optimized segmented pulses}

Here we apply a natural extension in the optimization protocol by using segmented pulses. This further broadens the searching parameter spaces, and the gate operation performance can be improved under the optimization protocols. This offers a multi-dimensional optimized shaped pulse that should be more robust to the pulse amplitude fluctuations. In the segmented pulse case, we still keep $\Omega_{r}$ and $\Delta_p$ constant in time but optimizable as in the previous subsection, and shape $\Omega_{p}$ by dividing it into $12$ symmetric segments with equal time periods $\Delta t$, which are connected by the error function (Erf) as \cite{Zahedinejad2016, Saffman2020},
\begin{eqnarray}
\Omega_{p}(t) =&& \frac{\Omega^{i+1}_{p,\text{max}}+\Omega^{i}_{p,\text{max}}}{2} \nonumber\\ 
&&+ \frac{\Omega^{i+1}_{p,\text{max}}-\Omega^{i}_{p,\text{max}}}{2}
{\rm Erf}\left[\frac{5}{\Delta t}\left(t-\frac{t^{i+1}+t^{i}}{2}\right)\right],
\end{eqnarray}
where $t_{i}\leq t\leq t_{i+1}$ for the $i$th segment. These segmented pulse amplitudes provide more optimization parameters and unexplored parameter regimes for potentially more improved $C_Z$ gates.  

\begin{figure}[t]
    \includegraphics[width=0.48\textwidth]{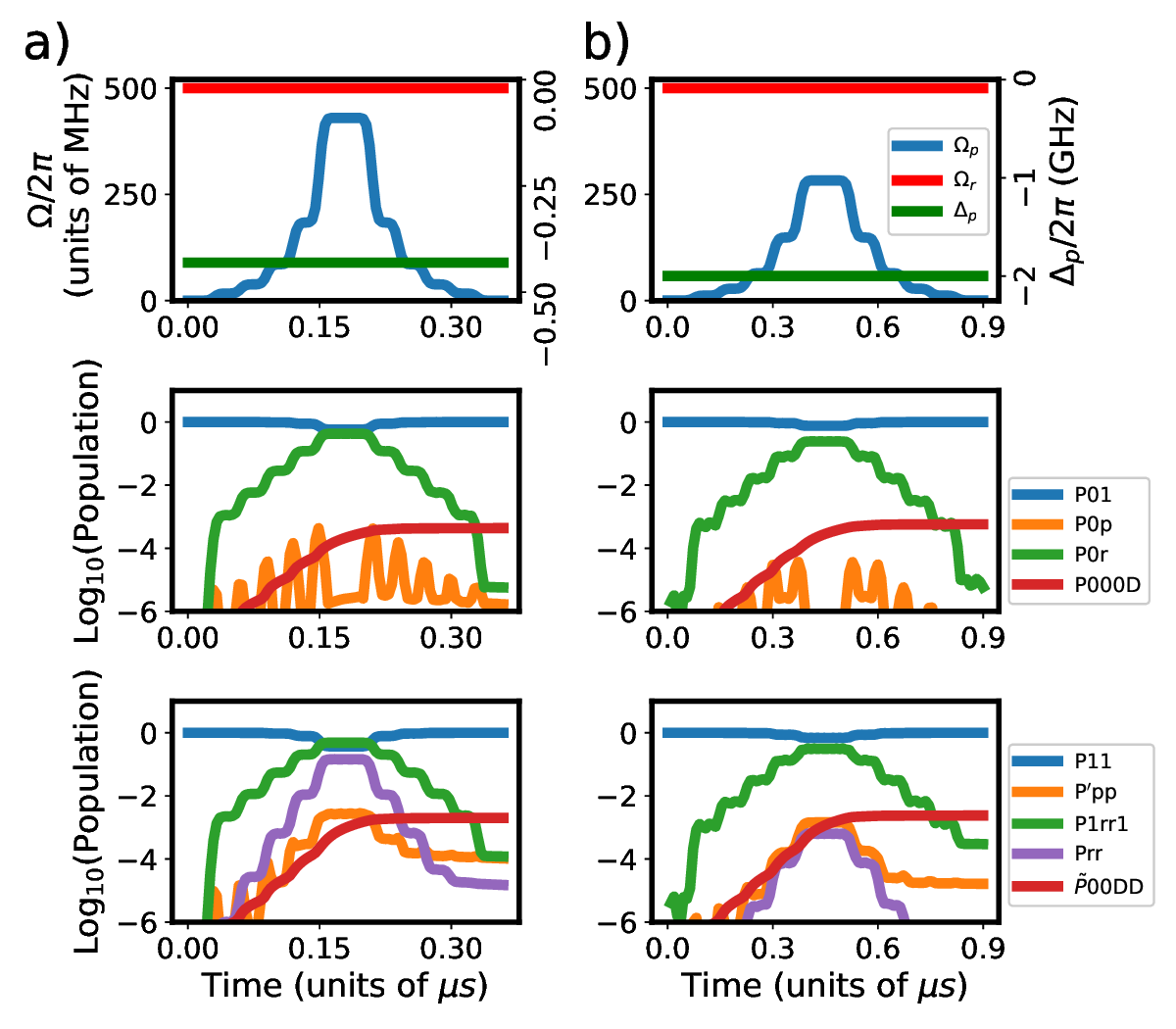}
    \caption{Optimized segmented pulses for two corresponding parameter regimes in Fig. \ref{fig3}. We again present in the top plots for the optimized pulse shape and laser parameters, in the middle and the bottom plots for the time evolutions of the atomic populations from the initial states $\ket {01}$ and $\ket{11}$, respectively. The range of total gate operation time and the ranges of the laser amplitudes are the same as in Fig. \ref{fig3}, where we obtain $F_{\rm Bell} \approx 0.99915$ and $0.99894$, respectively.}\label{fig4}
\end{figure}

Using the same parameters as in Fig. \ref{fig3}, we present the corresponding results in Fig. \ref{fig4} with segmented pulses as a comparison. The simulation results indicate that the $F_\text{Bell}$ and optimized $T_\text{gate}$ are similar to the optimized Gaussian pulses. Interestingly, the simulation results in Fig. \ref{fig4}(a) are in the nonadiabatic regime where the laser detuning is moderate comparing to the optimized value of $\Omega_r$ which hits again the boundary of the optimization range as in Fig. \ref{fig3}(a). This suggests a monotonic improvement of gate fidelity if the optimization range can be made even broader if there is no other physical limitation from the atomic system. Similar to the optimized Gaussian pulse scheme, the main error originates from the dark state population. The essential effect of optimized segmented pulses manifests in the continuing rises and falls of the atomic populations in the intermediate excited state $|0p\rangle$ from the initial state of $|01\rangle$. Therefore, as in the scheme using an optimized Gaussian pulse, the whole scheme using segmented pulses intends to suppress the populations in $|0p\rangle$, which is unavoidable in the two-photon excitation process.      

The example of using $12$ segments in Fig. \ref{fig4} is to demonstrate an enlarged optimization space to potentially improve the gate performance. As shown in the STIRAP pulse scheme \cite{Saffman2020}, a fidelity of $0.976$ using analytical pulse shapes can significantly be improved to $0.997$ with segmented pulses. In our case on the other hand, the best fidelity achievable using the segmented pulses is not much different from using an optimized Gaussian pulse presented in Sec. III A. Therefore, our demonstration in Fig. \ref{fig4} mainly offers a multi-dimensional optimized shaped pulse that should be more robust to the pulse amplitude fluctuations. This also suggests a remarkable success of the Gaussian pulse shape in the gate performance, which is more convenient and practical in experimental operations. 

We note that our simulation results are in huge contrast to the STIRAP regime \cite{Saffman2020}, where the initial states of either $|01\rangle$ or $|11\rangle$ almost completely transfers to the Rydberg states via the coupled qubit state $|1\rangle$. The results presented here indicate a new route to achieve a high-fidelity controlled-phase gate with initial states kept intact, where a genuine $C_Z$ gate operation can be realized in the nonadiabatic regime with required accumulated phases determined by the laser pulse areas. This nonadiabatic regime is successful in achieving a high-fidelity gate even with finite populations in the states $|r\rangle$ and $|p\rangle$. A finite population in $|r\rangle$ represents a weak blockade regime, which can still achieve high-fidelity gates under a high laser field. Meanwhile, the decoherence effect from $|p\rangle$ can be mitigated by applying a shorter and optimized pulse. 

\section{Laser fluctuation and Doppler dephasing}

Lastly, we investigate the effects from pulse amplitude fluctuations and Doppler dephasing on the gate performance. The dephasing on the laser detuning is due to the finite motions of the atoms under a finite temperature. At a temperature $T$, we assume both of the atoms moving with the root mean square velocity $v_{rms} = \sqrt{k_{\rm B} T/m}$ in the Maxwell-Boltzmann distribution along the copropagating direction of the laser beams, where $k_{\rm B}$ is the Boltzmann constant and $m$ is the mass of cesium atoms. The dephasing effect can be quantified as variations of the laser detunings $\delta \Delta_{p, (r)} = k_{p,(r)} v_{rms}$ with $k_p$ and $k_r$ the wave number of laser beams, respectively.

\begin{figure}[t]
    \includegraphics[width=0.48\textwidth]{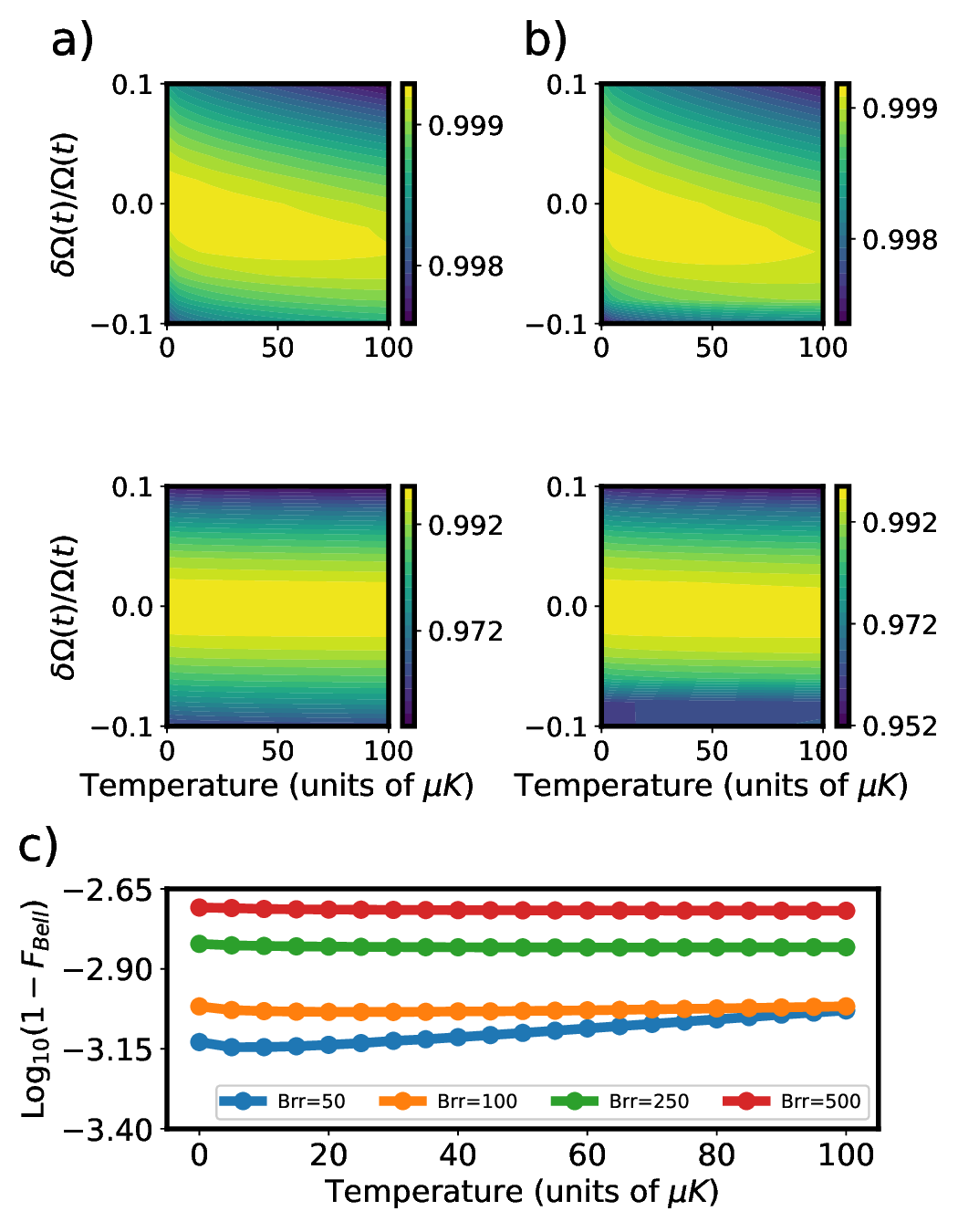}
    \caption{Bell fidelity under temperature effect and pulse amplitude fluctuations. The amplitude fluctuation is introduced as $\delta \Omega (t)$ in the $\Omega_p$ and $\Omega_r$ for the optimized a) Gaussian pulse and b) segmented pulse scheme at $B_{rr}/(2\pi)=50$ MHz and $\Delta_p/(2\pi)=-2$ GHz in the upper and lower plots, respectively, under the same system parameters in Figs. \ref{fig3} and \ref{fig4}. c) The influences of the temperature on several Bell infidelities from the horizontal cuts at $\delta\Omega(t)=0$ as in a) are shown for various $B_{rr}$.}\label{fig5}
\end{figure}

As shown in Fig. \ref{fig5}, we compare the variations of the optimized Gaussian and segmented pulses under $10\%$ laser field fluctuation and the temperature effect up to $100$ $\mu$K. The thermal fluctuations from Doppler dephasing modifies the laser detuning and the condition of two-photon resonance, to which the $C_Z$ gate performance is most robust. As the field fluctuates, the gate fidelity can be reduced by $4\%$ in the adiabatic regime of the bottom plot in Figs. \ref{fig5}(a) and \ref{fig5}(b). For the upper plots in the nonadiabatic regime, the variation of the fidelity is significantly reduced within $0.2\%$, as well as for different Rydberg blockade regimes in Fig. \ref{fig5}(c). This low sensitivity can be attributed to the high laser fields in the nonadiabatic regime, which seems to be promising in reaching an even higher fidelity and more robust controlled-phase gates but at the price of relatively high excitation powers.  

\section{Discussion and Conclusion}

We note that there should be more than two considered CZ gate errors from an intermediate state decay and an imperfect blockade of double Rydberg excitations. Some other errors can be relevant, for example, the lifetime of the Rydberg states, the atomic motional decoherence, uncertainty of Rydberg interaction strength owing to atomic position fluctuation, and unwanted populations in the nearby Rydberg states. Some of these additional errors can be made negligible if a cryogenic environment can be applied or a deeper potential in the optical tweezer trap can be utilized. These intrinsic and systematic errors would eventually become more involved and need to be addressed once a higher-fidelity gate operation is considered.            

Based on the gate performance in our optimization protocols using Gaussian or segmented pulses, we find that they both can realize similar high-fidelity and sub-microsecond $C_Z$ gates. This suggests that in optimized shaped pulses, there should exist multiple optimized solutions to achieve high-fidelity gate operations. To implement these gates experimentally, a convenient setup under amenable system parameters would be a practical consideration. Therefore, we focus on optimizing only one varied laser pulse shape with additional tunable time periods, while we keep the other laser and atomic system parameters constant in time but tunable in the optimization protocols. This consideration gives sufficient variables that can be optimized and as well be manageable in realistic implementations without many complications. Although Gaussian pulses seem to be adequate to implement a high-fidelity two-qubit gate, which are more easily generated and manipulated from an experimental perspective, the optimized segmented pulse scheme provides an alternative setup for high-fidelity gates that should be more resilient to environmental fluctuations or system uncertainties.    

In conclusion, we have theoretically studied a global laser driving scheme in a two-atom setup to achieve a high-fidelity $C_Z$ gate under the optimization protocols. With finite Rydberg interactions and the inclusion of intermediate excited state in the two-photon excitation setup, we numerically obtain the optimized Gaussian or segmented pulses in the symmetric gate operation protocol. Within a broad range of laser and atomic system parameters, the optimization protocol presents a suppressed excited-state and Rydberg-state population, which gives rise a $C_{Z}$ gate with a high Bell fidelity up to $0.9992$. These optimized pulses are robust to thermal fluctuations and the laser field variations. Our results promise an improved and fast gate operation under amenable and controllable experimental parameters. In particular, these optimized shaped pulses go beyond the adiabatic regime and suggest a high-fidelity gate performance under a weak Blockade regime. We have shown that the nonadiabatic operation regime can lead to improvements of the gate fidelity that the adiabatic regime cannot foresee and the potential of larger parameter spaces in optimization protocol can give rise to even better gate fidelities. Our results eventually provide new insights in generating a high-fidelity gate operation which is crucial in applications of deep quantum circuit design and fault-tolerant quantum computing.  

\section*{ACKNOWLEDGMENTS}
We acknowledge the support from the National Science and Technology Council (NSTC), Taiwan, under the Grant No. NSTC-109-2112-M-001-035-MY3, No. 112-2112-M-001-079-MY3, No. NSTC-
112-2119-M-001-007, and the support from Academia Sinica, Taiwan, under the project number AS-iMATE-110-36. We are also grateful for helpful discussions with Yi-Cheng Wang.  

\end{document}